# Spontaneous Hall effects in the electron system at the SmTiO$_3$/EuTiO$_3$ interface

**Kaveh Ahadi [†], Honggyu Kim, and Susanne Stemmer**

Materials Department, University of California, Santa Barbara, California 93106-5050, USA.

[†] Corresponding author.  Email: kahadi@mrl.ucsb.edu

**Abstract**

Magnetotransport and magnetism of epitaxial SmTiO$_3$/EuTiO$_3$ heterostructures grown by molecular beam epitaxy are investigated. It is shown that the polar discontinuity at the interface introduces ~ 3.9×10$^{14}$ cm$^{-2}$ carriers into the EuTiO$_3$. The itinerant carriers exhibit two distinct contributions to the spontaneous Hall effect. The anomalous Hall effect appears despite a very small magnetization, indicating a non-collinear spin structure and the second contribution resembles a topological Hall effect. Qualitative differences exist in the temperature dependence of both Hall effects when compared to uniformly doped EuTiO$_3$. In particular, the topological Hall effect contribution appears at higher temperatures and the anomalous Hall effect shows a sign change with temperature. The results suggest that interfaces can be used to tune topological phenomena in itinerant magnetic systems.

(Quasi-)two-dimensional electron systems at polar oxide interfaces have generated significant interest. For example, they allow for the study of phenomena associated with itinerant carriers in complex oxides without the introduction of dopant atoms [1]. Most studies thus far have focused on quasi-two-dimensional electron systems that reside in SrTiO$_3$ near the interface with another perovskite oxide containing nominally trivalent cations, such as LaAlO$_3$ [2] or Mott insulating $R$TiO$_3$ ($R$ is a rare earth ion but not Eu) [3]. For (001) interfaces, the $R^{3+}$O$^{2-}$ (Ti$^{3+}$O$_2^{4-}$) planes carry a +1 (-1) formal charges and encounter the charge-neutral layers of SrTiO$_3$ (Sr$^{2+}$O$^{2-}$ or Ti$^{4+}$O$_2^{4-}$). The very large (~$3.4\times10^{14}$ cm$^{-2}$) fixed charge at such interfaces is compensated by mobile electrons in the SrTiO$_3$.

Similar to SrTiO$_3$, EuTiO$_3$ is a band insulator with a $d^0$ electron configuration and has a non-polar (001) surface formed by alternating, charge neutral Eu$^{2+}$O$^{2-}$ and Ti$^{4+}$O$_2^{4-}$ planes. It exhibits G-type antiferromagnetism below the Neel temperature of 5.5 K [4]. With chemical doping, for example with a trivalent rare earth ion, it transitions to an itinerant (anti-)ferromagnet [5]. In addition, doped EuTiO$_3$ exhibits unique properties that directly reflect reciprocal and/or real space topologies. In particular, its anomalous Hall effect (AHE) changes sign with carrier density [6], which is an indication of regions in the Brillouin zone that carry a large Berry curvature near the Fermi level, such as (avoided) band crossings [7-9]. Non-collinear spin arrangements can produce a large contribution to the anomalous Hall effect even in antiferromagnets that have only a small net magnetization [10-12]. At low temperatures, an additional contribution appears in the Hall signal of doped EuTiO$_3$ films, which closely resembles the topological Hall effect (THE) [13]. The THE is caused by non-coplanar or chiral spin structures to which itinerant electrons couple [14, 15]. The carrier density dependence of these effects in EuTiO$_3$ suggests rich opportunities for controlling them in electrostatically doped heterostructures, which can also be used to gain further insights into the contributions to Hall effect in this material.

Here, we investigate the properties of polar/non-polar $SmTiO_3$/$EuTiO_3$ interfaces grown by molecular beam epitaxy (MBE). We show that the interface introduces mobile carriers into the $EuTiO_3$ and that carrier density closely matches the one expected from the polar discontinuity. The transport properties exhibit signatures of magnetic order and two distinct contributions to the spontaneous Hall effect, but there are qualitative differences compared to chemically doped $EuTiO_3$ films with similar carrier densities.

Heterostructures consisting of 10 pseudocubic unit cells (u.c.s) of $SmTiO_3$ on 60 nm thick $EuTiO_3$ epitaxial layers were grown by MBE on (001) $(La_{0.3}Sr_{0.7})(Al_{0.65}Ta_{0.35})O_3$ (LSAT) single crystals, as described elsewhere [13]. Cross-section scanning transmission electron microscopy showed abrupt interfaces (see Supplementary Information). Electron beam evaporation through a shadow mask was used to deposit Au/Ti (400/40 nm) contacts in square Van der Pauw configuration. Temperature-dependent magnetotransport measurements were carried out using a Quantum Design Physical Property Measurement System (PPMS). Magnetic properties were measured using a Quantum Design superconducting quantum interface device (SQUID) magnetometer.

Figure 1(a) compares the sheet resistances, $R_s$, as a function of temperature of a 60 nm-thick $EuTiO_3$ film and that of the $SmTiO_3$/$EuTiO_3$ heterostructure. While the $EuTiO_3$ film is highly resistive and quickly exceeds the measurement limit near room temperature, the heterostructure shows metallic behavior. This confirms that the $SmTiO_3$/$EuTiO_3$ interface is the source of mobile charge and metallic conductivity. An upturn in $R_s$ at ~30 K is followed by a relatively sharp peak at ~ 6 K. This upturn is similar to that found in uniformly doped $EuTiO_3$ films where it signifies the onset of magnetic order [13], confirming that the mobile carriers reside in the EuTiO3. Figure 1(b) shows the sheet carrier density ($n$) of the heterostructure, as determined from the ordinary Hall effect. At room temperature, $n = 3.9 \times 10^{14}$ cm$^{-2}$, which is close to the expected value ($3.4 \times 10^{14}$ cm$^{-2}$) from the polar discontinuity. The carrier density shows a weak, but noticeable, temperature dependence.

Figure 2(a) shows the magnetization of the heterostructure as a function of temperature under an in-plane applied magnetic field ($B$) of 100 Oe. Upon cooling, the magnetization increases around 6 K, consistent with the peak in the sheet resistance [Fig. 2(a)]. The remnant magnetization [Fig. 2(b)] is, however, very small. A superlinear increase in the magnetization appears near 0.3 T, hinting at a field-induced magnetic transition.

Figure 3(a) shows the longitudinal resistance, $R_{xx}$, as a function of out-of-plane magnetic field at different temperatures between 2 K and 30 K. The magnetic field was swept from +9 T to -9 T and back. Negative magnetoresistance is observed at all temperatures. Only a very small hysteresis exists at 2 K (see Supplementary Material). At 2 K, a sharp inflection can be noticed near 3 T. Figure 3(b) shows the Hall resistance at different temperatures, after antisymmetrizing and using a linear fit (6-9 T) to subtract the ordinary, linear Hall effect (the antisymmetrized raw data are shown in the Supplementary Information). In the presence of both AHE and THE, the Hall resistance, $R_{xy}$, is given as $R_0 B + R_{AHE} + R_{THE}$, where $R_0 B$ is the ordinary Hall component and $R_{AHE}$ and $R_{THE}$ are the AHE and THE contributions, respectively. At 30 K only the AHE is observed, seen in Fig. 3(b) as a monotonic change with $B$, which extrapolates to zero. Upon lowering the temperature, the AHE changes sign (is negative at 2 K and at 5 K) and an additional peak appears at 15 K, which is superimposed on the AHE. In contrast to the AHE, this additional, non-monotonic contribution does not change sign with temperature. The derivative of the longitudinal magnetoresistance with respect to the applied field (d$R_{xx}$/d$B$) shows a slope change at the field at which the peak in appears in the Hall effect [Fig. 3(c)].

Comparison of the results with previous studies of the Hall effect in chemically doped EuTiO$_3$ [6, 13] show that the two distinct contributions to the Hall effect are intrinsic properties of itinerant carriers in EuTiO$_3$ and do not depend on how these carriers are introduced, i.e. by electrostatic doping (this study) or by chemical doping (refs. [5, 6, 13]). Furthermore, both contributions to the Hall effect are observed despite an almost negligible net magnetization, consistent with their intrinsic, Berry phase origin. There are, however, qualitative differences in the properties of the interfacial electron system and uniformly doped EuTiO$_3$. In particular, the latter does not show a sign change in the AHE with temperature, only with doping [6]. The intrinsic AHE reflects the Berry curvature at the Fermi surface [7] and sign changes can occur near band crossings or other singularities in the density of states [7, 8, 11, 16]. The sign change of the AHE with temperature indicates that in the interfacial electron system the Fermi level passes through a singularity in the electronic density of states, where the Berry curvature changes sign, as the temperature is changed. The apparent decrease in carrier density with decreasing temperature [Fig. 1(b)] may thus indicate a change in the Fermi surface. Extrinsic effects, such as carrier trapping, are, however, also a possibility. The most likely origin for the differences between the heterostructures and the bulk lies in the differences in the electronic structure. Similar to SrTiO$_3$ [17, 18], we expect the interfacial carrier system to be not very strongly confined and to be better described as quasi-two-dimensional. Nevertheless, the confining potential of the fixed charge at the interface can have significant influence, for example, the orbital polarization [19].

Unlike the monotonic AHE, the second, non-monotonic contribution to the Hall effect does not change sign upon lowering the temperature. Furthermore, it appears at higher temperatures than in uniformly doped EuTiO$_3$, where it can only be detected at 2 K [13]. The fact that this contribution vanishes at high magnetic fields, when the spins become more collinear, indicates that it is likely related to a non-collinear spin texture. In particular, the results show that the non-collinear spin arrangement is stabilized to higher temperatures in the heterostructure than in the corresponding bulk materials. We note that the strict definition of the THE is that it is due to the real-space Berry curvature of chiral spin arrangements. As discussed in ref. [13], transport measurements alone are insufficient to definitely attribute the peaks seen in Fig. 3(b) to the THE. The results in this study, in particular the lack of a sign change, show, however, that it is a contribution that it is distinct from the monotonic AHE and furthermore that details of the spin texture are the origin of this contribution. This supports the previous assignment of the peaks to the THE.

In summary, the results emphasize the sensitivity of Hall effects in this materials system to the topology of the electronic states and spin textures, as expected from their Berry phase origin. Clearly, an improved understanding of the electronic structure and spin texture of the quasi-two-dimensional carrier system would be of great interest. Furthermore, the results show that the properties can be engineered using interfaces and heterostructures, which also open up other possibilities, such as electric field tuning of the carrier density.

See Supplementary Material for a scanning transmission electron microscopy image of a SmTiO$_3$/EuTiO$_3$ interface, a zoomed-in version of the magnetoresistance data at 2 K, and the raw Hall resistance data.

**Acknowledgments**

The authors thank Salva Salmani Rezaie for help with the TEM studies. We acknowledge support from the National Science Foundation under award no. ECCS 1740213. The microscopy studies were supported by the U.S. Department of Energy (Grant No. DEFG02-02ER45994). The work made use of the MRL Shared Experimental Facilities, which are supported by the MRSEC Program of the U.S. National Science Foundation under Award No. DMR 1720256.

**Figure Captions**

**Figure 1:** (a) Temperature dependence of $R_s$ of EuTiO$_3$(60 nm) with and without of a 10 u.c. SmTiO$_3$ cap layer, respectively. The inset shows the low temperature upturn in $R_s$ for the heterostructure. (b) Temperature dependence of the Hall carrier density of the SmTiO$_3$/EuTiO$_3$ heterostructure.

**Figure 2:** Magnetization of the SmTiO$_3$/EuTiO$_3$ heterostructure with in-plane applied magnetic field measured under field cooling (100 Oe). (b) Magnetization as a function of the in-plane applied field at 2 K.

**Figure 3:** (a) Longitudinal magnetoresistance ($R_{xx}$) of the SmTiO$_3$/EuTiO$_3$ heterostructure at different temperatures. (b) THE and AHE contributions to the Hall resistance ($R_{xy}$), obtained by subtracting the ordinary Hall effect, at different temperatures. (c) Applied out-of-plane magnetic field dependence of the derivative of the longitudinal resistance with respect to the magnetic field at 2K.

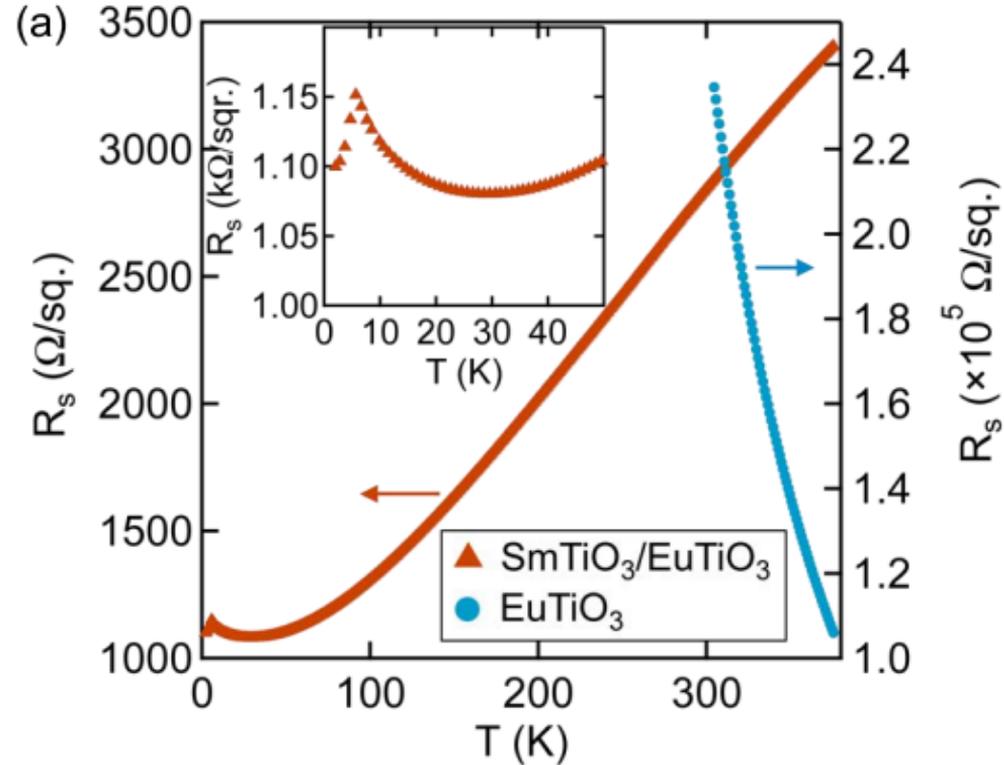

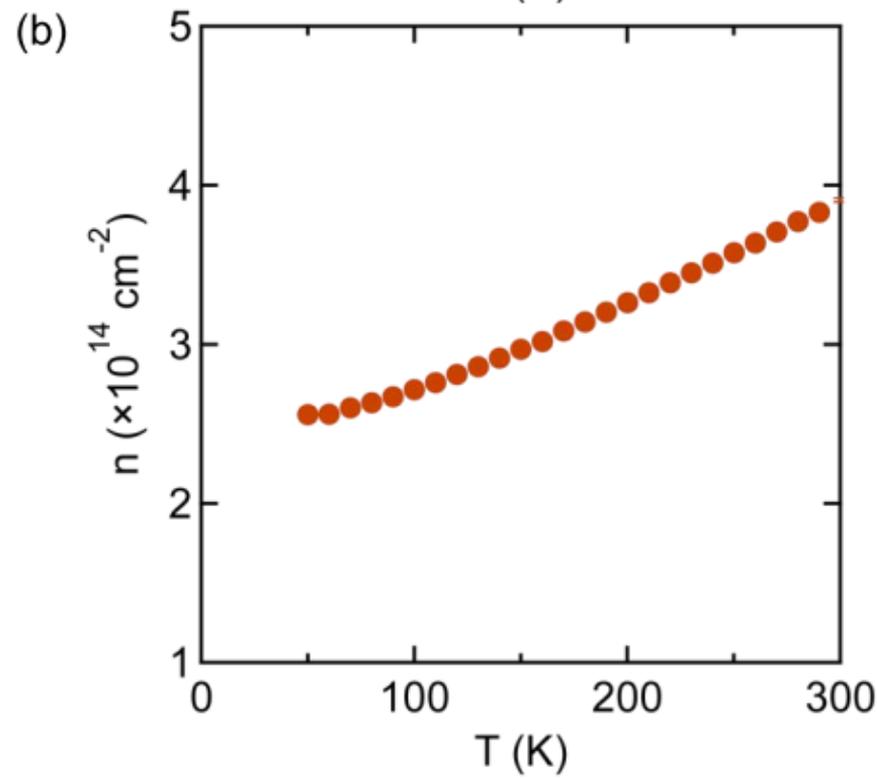

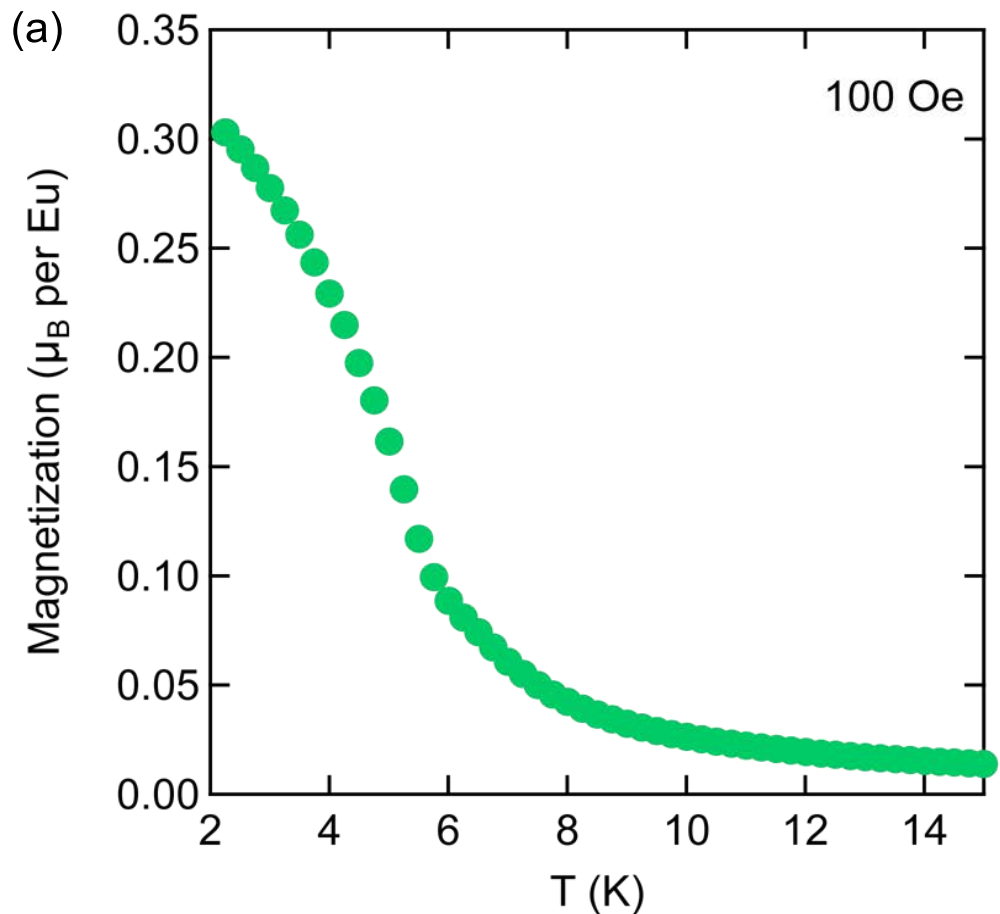

(a)

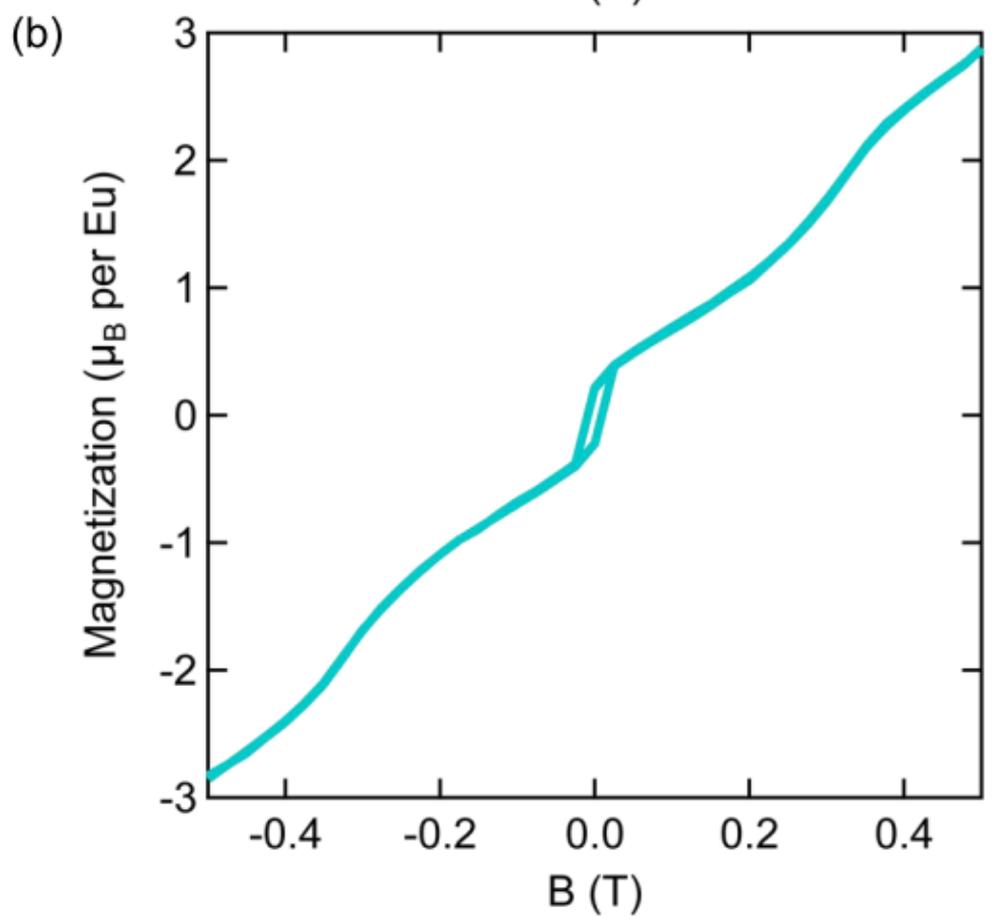

(b)

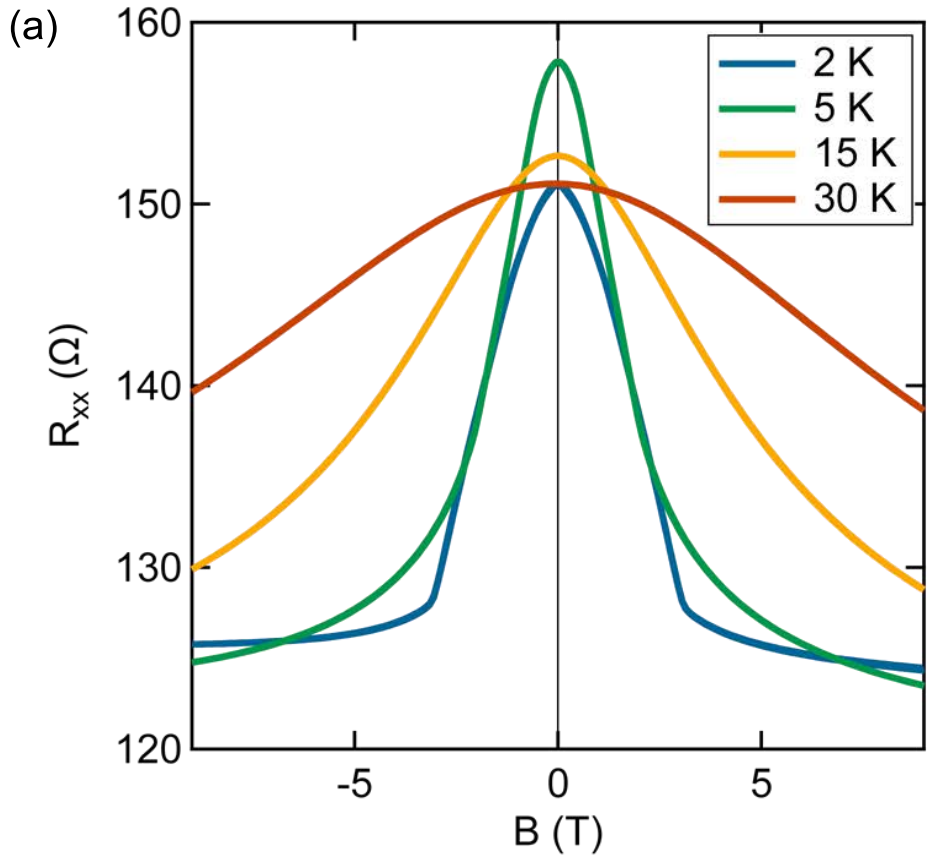

(a)

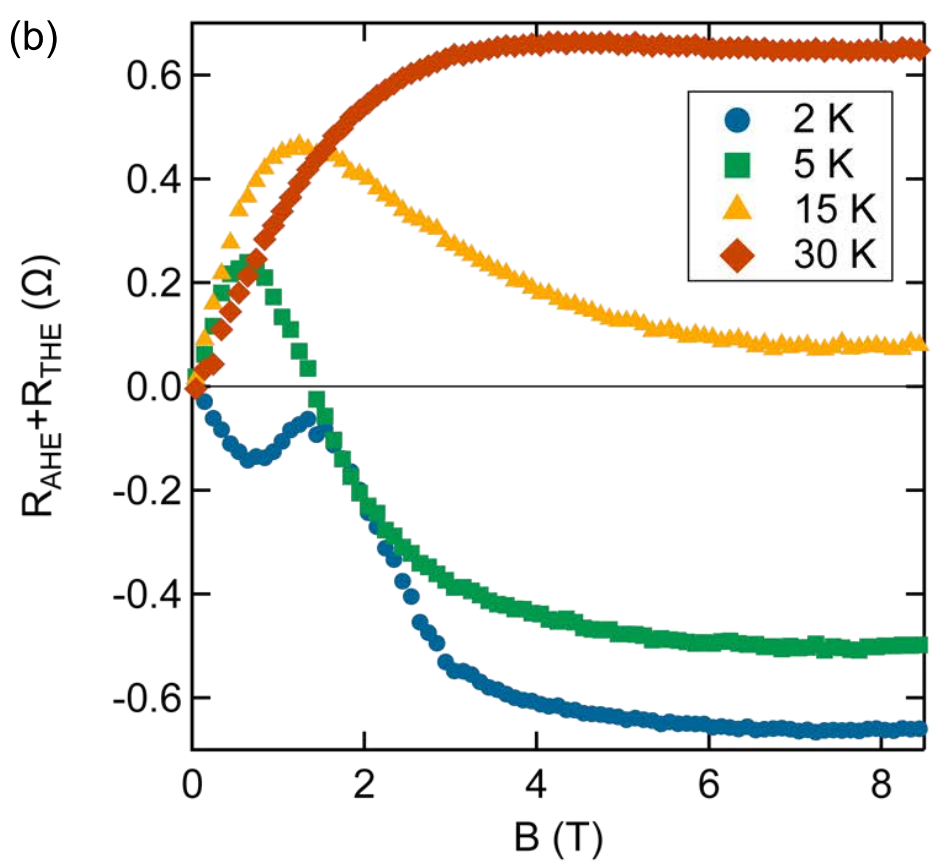

(b)

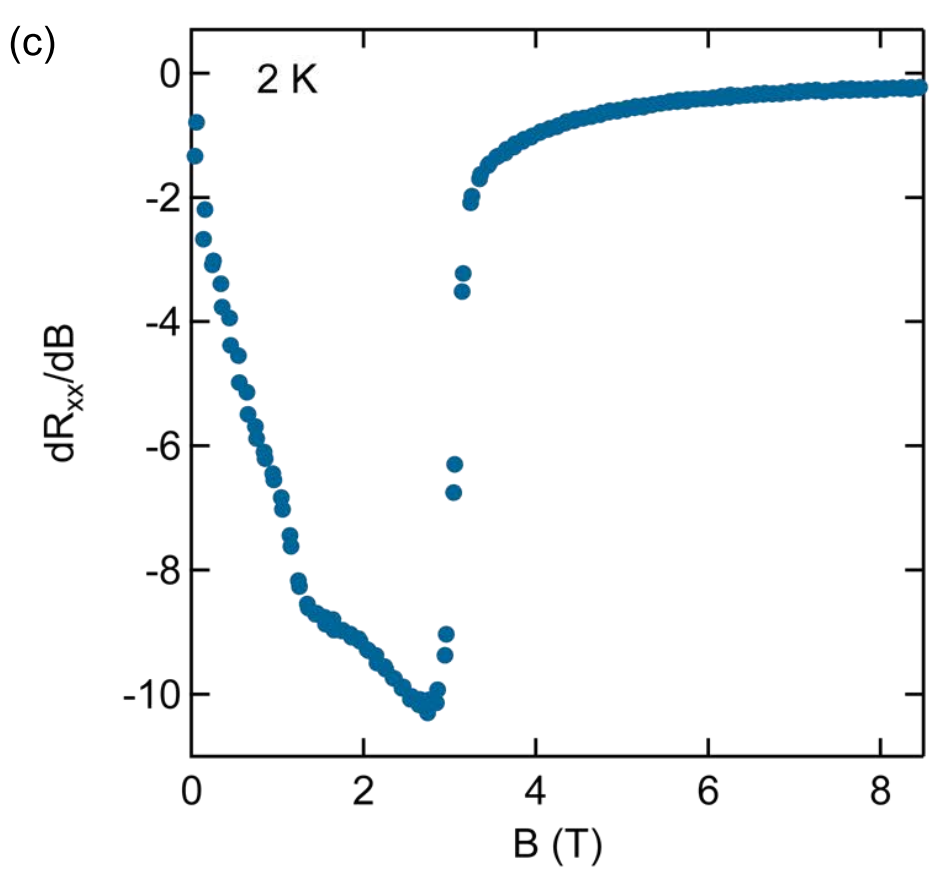

(c)